\documentclass[aps,reprint,superscriptaddress,longbibliography]{revtex4-1}

\usepackage{setspace}

\usepackage{tikz, calc}
\usepackage{amsmath}
\usetikzlibrary{shapes.geometric, arrows.meta, positioning}

\usepackage{hyperref}
\usepackage{outlines}
\usepackage{fancyvrb}
\usepackage{amsmath}
\usepackage{hyperref}
\usepackage{braket}
\usepackage{comment}
\usepackage{bbold}
\usepackage{amssymb}
\usepackage{graphicx}
\usepackage[caption=false]{subfig}
\usepackage{soul}
\usepackage{amsthm}
\usepackage{qcircuit}

\usepackage{yquant}
\usepackage{algorithm}      
\usepackage{algpseudocode}  

\yquantdefinebox{circle}[shape=yquant-circle, draw,
inner sep=0pt, radius=0.5mm]{}

\begin{document}

\title{Photonic implementation of quantum hidden subgroup database compression}



\author{Qianyi Wang}
\thanks{These authors contributed equally to this work.}
\affiliation{School of Physics, College of Engineering and Applied Sciences, Nanjing University, Nanjing, 210093, China}

\author{Feiyang Liu}
\thanks{These authors contributed equally to this work.}
\affiliation{Department of Physics, City University of Hong Kong, Tat Chee Avenue, Kowloon, Hong Kong SAR, China}

\author{Teng Hu}
\thanks{These authors contributed equally to this work.}
\affiliation{School of Physics, College of Engineering and Applied Sciences, Nanjing University, Nanjing, 210093, China}

\author{Kwok Ho Wan}
   \affiliation{Blackett Laboratory,
  Imperial College London, London, SW7 2AZ, United Kingdom}

\author{Jie Xie}
\affiliation{School of Physics, College of Engineering and Applied Sciences, Nanjing University, Nanjing, 210093, China}

\author{M.S. Kim}
  \affiliation{Blackett Laboratory,
  Imperial College London, London, SW7 2AZ, United Kingdom}

\author{Huangqiuchen Wang}
\affiliation{School of Physics, College of Engineering and Applied Sciences, Nanjing University, Nanjing, 210093, China}

\author{Lijian Zhang}
\thanks{Lijian Zhang \\(lijian.zhang@nju.edu.cn)}
\affiliation{School of Physics, College of Engineering and Applied Sciences, Nanjing University, Nanjing, 210093, China}

\author{Oscar Dahlsten }
\thanks{Oscar Dahlsten \\(oscar.dahlsten@cityu.edu.cn)}
\affiliation{Department of Physics, City University of Hong Kong, Tat Chee Avenue, Kowloon, Hong Kong SAR, China}
\affiliation{Institute of Nanoscience and Applications, Southern University of Science and Technology, Shenzhen 518055, China}



\date{\today}

\begin{abstract}
We experimentally demonstrate quantum data compression exploiting hidden subgroup symmetries using a photonic quantum processor. Classical databases containing generalized periodicities—symmetries that are in the worst cases inefficient for known classical algorithms to detect—can be efficiently compressed by quantum hidden subgroup algorithms. We implement a variational quantum autoencoder that autonomously learns both the symmetry type (e.g., $\mathbb{Z}_2 \times \mathbb{Z}_2$ vs. $\mathbb{Z}_4$) and the generalized period from structured data. The system uses single photons encoded in path, polarization, and time-bin degrees of freedom, with electronically controlled waveplates enabling tunable quantum gates. Training via gradient descent successfully identifies the hidden symmetry structure, achieving compression by eliminating redundant database entries. We demonstrate two circuit ansatzes: a parametrized generalized Fourier transform and a less-restricted architecture for Simon's symmetry. Both converge successfully, with the cost function approaching zero as training proceeds. These results provide experimental proof-of-principle that photonic quantum computers can compress classical databases by learning symmetries inaccessible to known efficient classical methods, opening pathways for quantum-enhanced information processing.
\end{abstract}

\maketitle

\newpage


{{\bf\em General introduction.---}}The quantum hidden subgroup algorithm refers to a family of quantum algorithms that find symmetries of some black box function $f(x)$, transformations on the input $x$ that leave the output $f(x)$ invariant. A famous example is the quantum period finding part of Shor's factoring algorithm, which is exponentially faster than known classical algorithms. More generally, this family encompasses a broad class of problems for which no efficient classical algorithms are known, but efficient quantum algorithms often exist~\cite{kaplan2016breaking,wang2021quantum,ben2020symmetries,kitaev1995quantum,ettinger2004quantum,lomont2004hidden,jozsa2001quantum, nielsen2002quantum}.

The quantum hidden subgroup algorithm has applications in data compression. Compressing data means reducing the number of logical bits used to store or transmit the data~\cite{goldberg1991compression}. The quantum hidden subgroup algorithm may extract features of the classical data that are not efficiently accessible to classical computers with known methods~\cite{liu2024information}. Those features can then be combined with an autoencoder structure to achieve information compression~\cite{liu2024information}.
\begin{figure}[h]
\centering
\includegraphics[width=\linewidth]{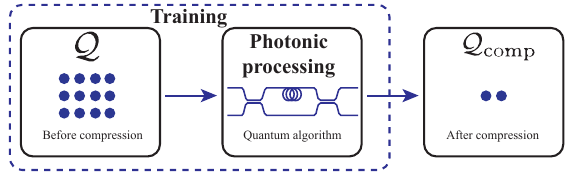}
\caption{{\bf Schematic representation of database compression through photonic quantum hidden subgroup compression.} The initial database  $\mathcal{Q}$ has duplication of entries in accordance with a hidden subgroup symmetry. $\mathcal{Q}$ undergoes interaction with a quantum photonic system via a quantum hidden subgroup algorithm to find generalized periods, associated with repeated entries in $\mathcal{Q}$. Training is used to find the symmetry the database has and this results in a compressed database $\mathcal{Q}_{\text{comp}}$.}
\label{fig:database_compression}
\end{figure}

Autoencoders, automatic encoders, are feedforward neural networks that can be trained on a given source to reversibly compress data from that source~\cite{ackley1985learning,kingma2019introduction}. They typically have a bottleneck layer in the middle. The neural network is trained, with that bottleneck constraint in place, on a given source to ensure that the input is reproduced at the final layer. Upon successful training of the network, the subnetworks before/after the bottleneck then compress/decompress the data respectively. The part before the bottleneck is called the encoder and the part after the decoder. 

A {\em quantum} version of autoencoders, proposed independently in Refs.~\cite{wan2017quantum, romero2017quantum}, generalizes classical autoencoders in a similar way to how quantum computing generalizes classical computing. Quantum autoencoders have been applied to encode~\cite{bravo2021quantum}, compress~\cite{wan2017quantum, romero2017quantum,cao2021noise}, classify and denoise quantum data~\cite{bondarenko2020quantum} and have been implemented experimentally~\cite{zhou2022preserving,   huang2020realization, zhang2022resource, ding2019experimental,  pepper2019experimental}. 

{\em Quantum hidden subgroup} autoencoders adapt the type of circuits used in quantum hidden subgroup algorithms as part of the encoder, to enable efficient compression of data with hidden subgroup symmetries, that known classical methods cannot compress efficiently~\cite{liu2024information}. Variational training of a generalized circuit ansatz can identify which variant of hidden subgroup symmetry is present in the data, e.g., Shor vs. Simon's symmetry, without prior knowledge~\cite{liu2024information}.

We here provide experimental demonstration of quantum hidden subgroup autoencoder database information compression on a quantum photonic platform.  Quantum photonics is one of the main candidate hardware platforms for quantum information processing~\cite{Geoff2019applied_physics_review,Kok2007RevModPhys,Knill2001Nature}, exploiting quantum states encoded in multiple degrees of freedom associated with light~\cite{Li25,Erhard2018light,Dada2011Nature_physics,Roslund2014NP,Kues2017Nature,Humphreys2013PRL,Brendel1999PRL,Brecht2015PRX,Humpreys2013PRL}. 
As outlined in Fig.~\ref{fig:database_compression}, we train a quantum photonic system with tunable gates on a given data source. The system involves the spatial, temporal and polarization modes of a single photon. We employ different ansatzes for the circuit architecture within which the training is performed. One ansatz closely follows the generalized Fourier transform family of circuits and one ansatz is more general in terms of the single qubit gates allowed. We find that training is successful for both ansatzes. We thus demonstrate experimental proof of principle of a method that leverages the hidden subgroup symmetry to achieve compression of classical databases that is in general inefficient for known classical approaches. 

{{\bf\em Quantum Hidden subgroup algorithm.---}}
The quantum hidden subgroup algorithm~\cite{ettinger2004quantum, lomont2004hidden, jozsa2001quantum}
 finds a generalized period of a function $f(x)$. By generalized period we mean the smallest non-trivial $r$ satisfying  
 \begin{equation}
 \label{eq:f}
 f(x)=f(x +_g r)\quad \forall x,
 \end{equation}
 where $+_g$ is the addition operation of the group in question. For example,
 for (the essential variant of) Simon's problem, $x$ and $r$ are bit strings with $r$ often denoted as $s$. $+_g$ corresponds to bitwise addition modulo 2, denoted by `$\oplus$'. Thus Eq. \ref{eq:f} becomes $f(x)=f(x\oplus s)$. $x$ and $s$ are then part of the group $\mathbb{Z}_2^n$, the n-fold Cartesian product of $\mathbb{Z}_2$, the group of integers with addition modulo 2, with itself. Commonly, one considers finite and abelian groups. $r$ then generates a finite subgroup by taking powers of $r$, which is the subgroup one wishes to find.  
 
 The essential idea of the quantum hidden subgroup algorithm~\cite{jozsa2001quantum} is to (i) generate the `quantum data-table' state  $\sum_{x}\ket{x}_{IN}\ket{f(x)}_{OUT}$, where the subscripts IN and OUT denote the qubits associated with the input ($x$) and output ($f(x)$) respectively; (ii) measure the OUT register, collapsing the state to $(\ket{x_0}_{IN}+\ket{x_0+r}_{IN}+\ket{x_0+2r}_{IN}+...)\ket{f(x_0)}_{OUT}$ where $\ket{f(x_0)}_{OUT}$ is determined by the random measurement outcome. (A physical measurement of the OUT register is not essential--tracing out OUT yields a mixture of such collapsed states on IN.) $x_0$ is an undesired uncontrolled overall shift making it difficult to measure 
 $r$; (iii) apply the (often inverse) generalized quantum Fourier transform on the IN register to move that $x_0$ shift into the phase of each term so that $x_0$ does not impact the probability of measurements in the computational basis on the IN register; (iv) measure IN in the computational basis; (v) repeat the procedure to build up data, and then analyse the data to infer $r$ of Eq. \ref{eq:f} to a high probability. 

Some comments on steps (i) and (iii). Firstly, step (i) is, in general, non-trivial because it involves designing a unitary $U_f$ for which 
 \begin{equation}
  \label{eq:U_f}
U_f\ket{x}_{IN}\ket{0}_{OUT}=\ket{x}_{IN}\ket{f(x)}_{OUT}.
 \end{equation}
The number of uses of $U_f$ of Eq. \ref{eq:U_f}, the  query complexity, does not count the cost of creating $U_f$ itself. Secondly, the unitary after $U_f$, the generalized Fourier transform in step (iii) is a tensor product of standard Fourier transforms on subsets of qubits~\cite{ettinger2004quantum, lomont2004hidden, jozsa2001quantum}. The standard quantum Fourier transform takes computational basis states $\ket{j} \mapsto \sum_{k=0}^{N-1}\frac{1}{\sqrt{N}}\omega^{jk}_{N}\ket{k}$, where $\omega_{N} = e^{\frac{2\pi i}{N}}$. The standard quantum Fourier transforms, and thus tensor products thereof, can be implemented efficiently via Hadamards and 2-qubit controlled rotation gates~\cite{ettinger2004quantum, lomont2004hidden, jozsa2001quantum} (See Supplemental Material for more details).

\begin{figure*}[ht]
\includegraphics[width=\linewidth]{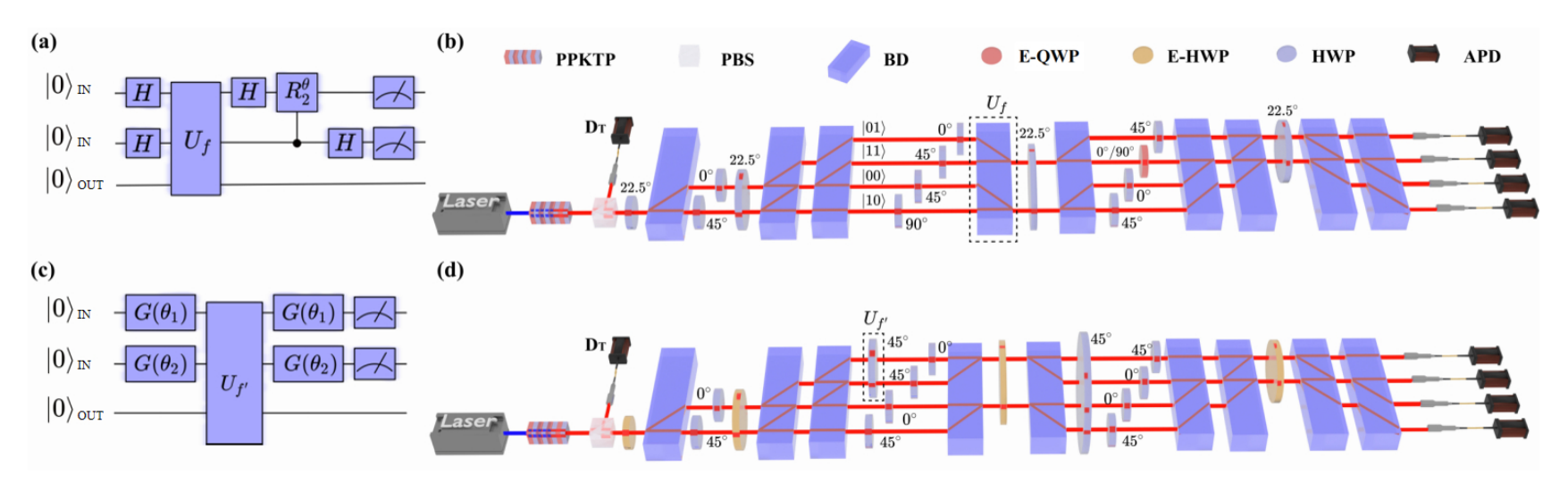}
\caption{\label{fig:exp_up_all}{\bf Photonic set-up for quantum part of information compression via tuneable quantum hidden subgroup algorithm.}
(a) Quantum circuit for information compression. The parameter $\theta$ governs the controlled gate (when $\theta=1$, the gate is active, and when $\theta=0$, it is inactive). The top two qubits are the IN-register associated with 
$x$, and the bottom qubit is the OUT-register associated with $f(x)$.
(b) Experimental setup for database compression. Pairs of photons are generated via spontaneous parametric down-conversion (SPDC) by pumping a periodically poled KTiOPO$_4$ (PPKTP) crystal. One photon is immediately detected by an avalanche photodiode (APD), $\mathrm{D}_\mathrm{T}$, to serve as a trigger, while its twin is directed into the subsequent setup. From bottom to top, the four spatial paths encode the IN-register qubit states $\ket{10}_{IN}$, $\ket{00}_{IN}$, $\ket{11}_{IN}$, and $\ket{01}_{IN}$, respectively. A beam displacer (BD) implements the oracle unitary operation $U_f$, and an electrically controlled quarter-wave plate (E-QWP) serves as the switch for the controlled gate. In this configuration, setting the E-QWP at $90^\circ$ leaves the IN qubits unchanged, while setting it at $0^\circ$ applies a phase shift of $e^{i\pi/2}$ specifically to photons in the $\ket{11}$ state. A polarizing beam splitter (PBS) and half-wave plates (HWPs) are also used to modulate states of photons.
(c) Tuneable circuit capable of finding Simon's symmetry. $\theta_1$ and $\theta_2$ are tuneable parameters.
(d) Experimental setup for finding Simon's symmetry in the case $n=2$. Tunable gates shown in (c) are realized by four electronically controlled half-wave plates (E-HWPs). The oracle unitary operation is applied by a fixed HWP. From bottom to top, the four path modes, correspond to $\ket{00}_{IN}$, $\ket{01}_{IN}$, $\ket{10}_{IN}$, and $\ket{11}_{IN}$ for $s=01$, and to $\ket{00}_{IN}$, $\ket{10}_{IN}$, $\ket{01}_{IN}$, and $\ket{11}_{IN}$ for $s=10$, respectively.}
\end{figure*}

{{\bf\em Quantum hidden subgroup database compression.---}} Quantum hidden subgroup database compression can be defined as follows~\cite{liu2024information}. There is a classical, non-quantum,  database, denoted as $\mathcal{Q}$, which stores data. For concreteness one may think of this data as time-series data $f(x)$, where 
$x$ is the (discrete) time and $f(x)$ the value of the time series at time $x$. We shall represent $x$ and $f(x)$ as classical bit strings. The database $\mathcal{Q}$ is an ordered list of $f(x)$ entries, $\{f(x_1),f(x_2),...\}$. 
Compressing the database means to reduce the number of entries. Here, the compression will be done via de-duplication, removing {\em duplicated entries}, to create another, compressed database $\mathcal{Q}_{comp}$ with fewer entries, together with a record of which entries were duplicated, such that one may find $f(x)$ for any $x$ given access to $\mathcal{Q}_{comp}$. We shall suppose that the generating function $f(x)$ has a hidden subgroup symmetry as per Eq. \ref{eq:f}. Then the hidden subgroup symmetry imposes a hidden pattern in the time series data generated by the function $f$, which is a redundancy that can be eliminated to achieve data compression. 

The type of symmetry underlying the duplication of data is learned. We do not know in advance what circuits we should use to implement the quantum hidden subgroup algorithm. Instead, we use a tunable circuit to variationally find the suitable circuit, in particular the suitable generalized Fourier transform. Details about the variational circuit ansatz are given below (and in the Supplementary material). The tuning step is done within an autoencoder structure, such that the system is trained to send only information about the hidden subgroup and the generalized period through the bottleneck, onto a classical decoder (See the Supplementary Material for details).

{{\bf\em The database and its hidden symmetry.---}}We consider a toy database designed to have a hidden subgroup symmetry. The database consists of four data points corresponding to four sequentially indexed values of $x:\{ 00, 01,10,11 \}$ (recall that $x$ may be thought of as the time and $f(x)$ as the value of the time series at that time). 

Denoting two bit strings as $a_1a_2$ and $b_1b_2$, the group addition can be define in two distinct ways of interest here:

I. The group addition can be defined as 
\begin{equation}
a_1a_2 +_g b_1b_2=a_1a_2 \oplus b_1b_2, 
\label{eq:abxy1}
\end{equation}
where $\oplus$ signifies bit-wise addition modulo 2. Under this definition, the group $G$ is isomorphic to  $\mathbb{Z}_{2} \times \mathbb{Z}_{2}$.

II. The  group addition can be defined as
\begin{equation}
a_1a_2 +_g b_1b_2=a_1a_2 + b_1b_2 \mod 4,
\label{eq:abxy2}
\end{equation}
where $+$ is the standard arithmetic addition operation. Under Eq. \ref{eq:abxy2}, $G$ is isomorphic to $\mathbb{Z}_{4}$. 

We use, as will be described later, an experimental circuit ansatz that is capable of adapting to both cases I and II, and choose a database with symmetry of type I to demonstrate the training. In particular, we shall consider a database-generating function, $f$, such that $f(00)=f(01)=1$ and $f(10)=f(11)=0$, which generates a nontrivial four-point database with a unique hidden subgroup symmetry. Thus, the group of our toy database is isomorphic to  $\mathbb{Z}_{2} \times \mathbb{Z}_{2}$ and the generalized period can be seen to be $s=01$.

{{\bf\em The circuit ansatz.---}}To automatically uncover the hidden subgroup symmetry of the database, we have developed a variational autoencoder that comprises a quantum encoder and a classical decoder. We here focus on describing the quantum circuit for the {\em encoder}, which is illustrated in Fig. \ref{fig:exp_up_all}(a). (See the Supplementary Material for the full autoencoder structure)

The unitary operator $U_f$ acts on four qubits. $U_f$, defined in Eq. \ref{eq:U_f}, 
writes $f(x)$ onto the OUT register when given $x$ on the IN register as input. 
In the circuit of Fig. \ref{fig:exp_up_all}(a), the top two qubits serve as the IN-register qubits and the bottom qubit functions as the OUT-register.

The operator $CR_2^{\theta}$ implements a controlled phase gate. By definition, 
\begin{equation}
\label{eq:controlledphasetheta}
CR_2^{\theta}=\ket{0}\bra{0}\otimes I+\ket{1}\bra{1}\otimes R_2^{\theta}, 
\end{equation}
where $I$ is identity  and $R_2^{\theta}=|0\rangle\langle0|+e^{i\theta \pi/2}|1\rangle\langle 1|$. $\theta$ is a discrete parameter. When $\theta=0$, the controlled gate acts as the identity; when $\theta=1$, it introduces a phase shift of $e^{i\pi/2}$ to the $\ket{11}$ state.

By optimizing the encoder, we can infer the hidden subgroup symmetry underlying the function $f$.
For $\theta=0$, the circuit applies individual quantum Fourier transforms (QFTs) to each IN qubit, enabling identification of the hidden subgroup for functions defined on $\mathbb{Z}_2 \times \mathbb{Z}_2$. In contrast, for $\theta=1$, by classically permuting the measurement outcomes of the IN qubits at the end, the circuit implements a two-qubit QFT, thereby allowing for the identification of the hidden subgroup for a function defined on the group $\mathbb{Z}_4$. Based on the symmetry information provided by the encoder after training—including the group addition operation and the generalized period—we can reconstruct the database by assigning the same function value to all bit strings that differ by the generalized period.

{{\bf\em The photonic implementation.---}}We implement the quantum circuit depicted in Fig.~\ref{fig:exp_up_all}(a) utilizing a photonic platform, with the corresponding experimental setup illustrated in Fig.~\ref{fig:exp_up_all}(b).  

The single-photons are generated via a  spontaneous parametric down-conversion (SPDC) process. A diode laser (MOGbox DLC202) with a central wavelength of 405 nm is used to pump a periodically poled KTiOPO$_4$ (PPKTP) crystal to generate pairs of photons. The detection of one of the pair by the avalanche photodiode (APD) denoted as $\mathrm{D}_\mathrm{T}$ acts as a herald for the presence of the other photon.

The two IN qubits are encoded in the four spatial paths of a single photon, arranged (from bottom to top) according to the computational basis $\ket{10}$, $\ket{00}$, $\ket{11}$, and $\ket{01}$. Meanwhile, the OUT qubit is encoded in two distinct time bins, denoted by $\ket{0}$ and $\ket{1}$. 

We apply unitary gates to an individual IN qubit by converting its path encoding into polarization encoding using beam displacers (BDs) and wave plates. When photons pass through a BD, horizontally polarized ones are displaced to a different path, while vertically polarized photons remain in the the original path. As a result, photons originally in two different paths—each with a distinct polarization—can be made to coincide in the same path but with different polarizations. In this way, we effectively transfer path-encoded states into polarization-encoded states. Subsequently, polarization rotations are performed using half-wave plates (HWPs) and quarter-wave plates (QWPs). Their Jones Matrices are given in the Supplemental Material. After the polarization manipulations, the qubits are converted back to path encoding using another set of BDs. This process enables the implementation of the desired quantum gates on the IN qubits encoded in path modes. 

The oracle unitary $U_f$ is implemented by a BD that delays the $\ket{00}$ and $\ket{01}$ paths relative to $\ket{10}$ and $\ket{11}$. We identify the early time‐bin with $\ket{0}$ (encoding $f=0$) and the delayed time‐bin with $\ket{1}$ (encoding $f=1$). This configuration effectively encodes the function values in the photonic time-bin degree of freedom.

The controlled phase shift gate illustrated in Fig.~\ref{fig:exp_up_all}(a) is implemented using an electrically controlled quarter-wave plate (E-QWP). As depicted in Fig.~\ref{fig:exp_up_all}(b), the E-QWP is placed exclusively in the optical path corresponding to the IN qubit state $\ket{11}$. Photons in this path are prepared in the vertically polarized state before traversing the E-QWP, whose effect is determined by the orientation of its optical axis, $\theta_q$, relative to the horizontal polarization axis. When $\theta_q = 90^\circ$, the E-QWP applies the transformation $U_{\text{QWP}}(90^\circ)$, which leaves $\ket{11}$ unchanged, effectively switching off the controlled phase shift gate. Conversely, when $\theta_q = 0^\circ$, the E-QWP induces a phase shift of $e^{i\pi/2}$ to the $\ket{11}$ state, thus turning on the controlled phase shift gate. By electronically switching the optical axis angle of the E-QWP between $0^\circ$ and $90^\circ$, we can effectively control the activation of the controlled phase shift gate in Eq. \ref{eq:controlledphasetheta}. 

\begin{figure}[ht]
\includegraphics[width=\linewidth]{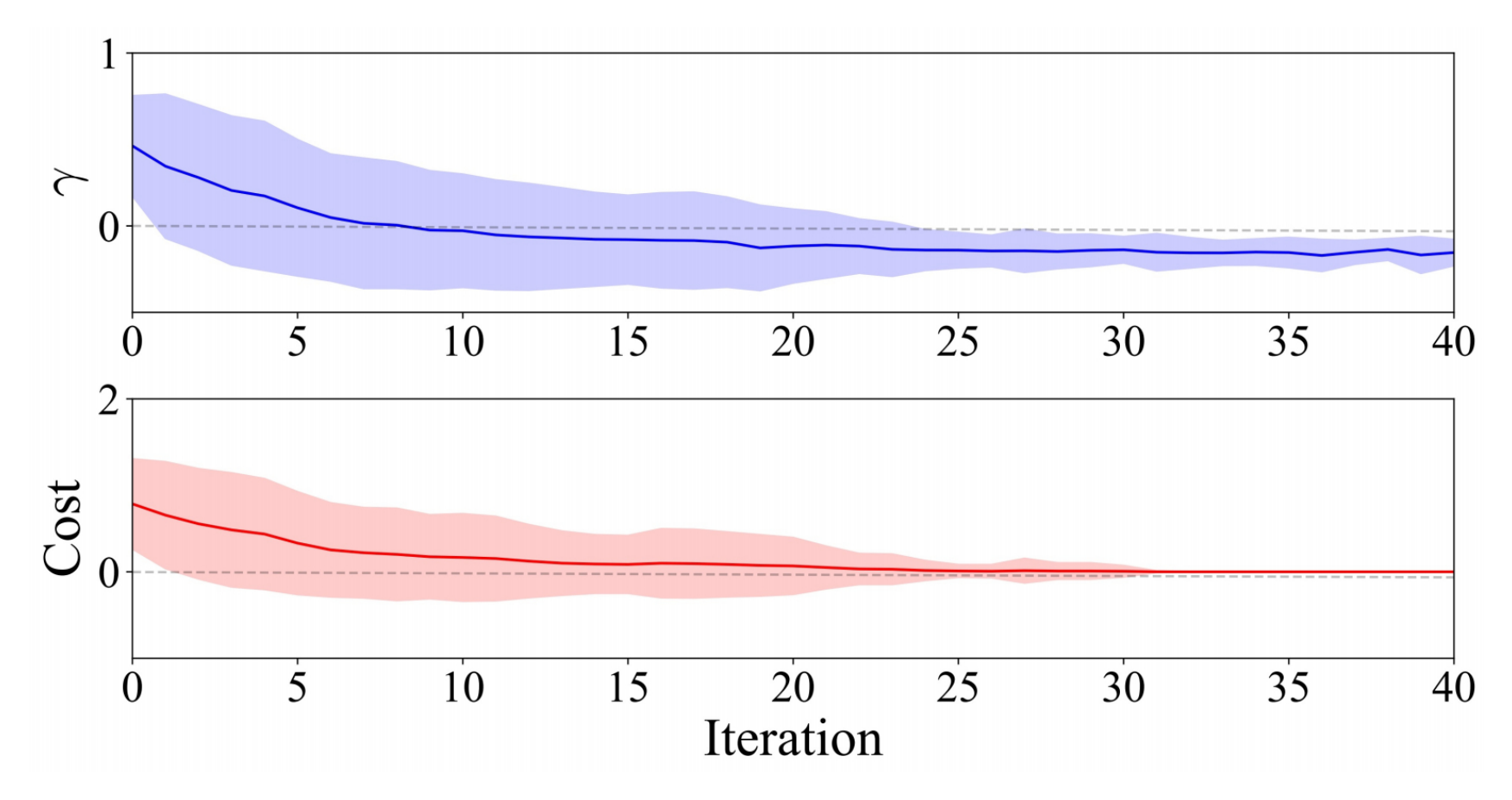}
\caption{\label{fig:Exp_result_compression} {\bf Training succeeds in finding the symmetry.}  $\gamma$ is a tuneable parameter such that $p(\theta=1)=\mathrm{max}(0,\mathrm{min}(\gamma,1))$, where $\theta$ is shown in Fig.~\ref{fig:exp_up_all} (a). Solid lines represent the mean values, while shaded areas denote values within one standard deviation of the mean.}
\end{figure}

{{\bf\em Experimental Results.---}} The results of the variational training process are shown in Fig. ~\ref{fig:Exp_result_compression}.
The figure shows the evolution of the cost function, the number that characterises how badly the given parameters achieve the task at hand,  as well as the parameter $\gamma$ over 40 iterations, where $\gamma$ controls the probability that $\theta=1$. This probability is given by the clamp function: $g(\gamma) = \text{max}(0,\text{min}(\gamma,1))$, so that $p(\theta=1)=g(\gamma)$.  The lower (red) plot illustrates the behavior of the cost function: 
\begin{equation}
    C(\gamma) = \sum_i(f(i)-\hat{f}(i))^2
\end{equation}
which quantifies the difference between the recovered database entry $\hat{f}(i)$ and the original database entry $f(i)$. Details about the measurement of this cost function and the training of the autoencoder can be found in the Supplementary material. As the training proceeds via gradient descent, the cost function drops steadily and eventually converges to zero—a clear indication of successful training. Furthermore, the shaded area, representing one standard deviation from the mean, significantly narrows as the training converges. This reduction in variance highlights the robustness and stability of the learning process.
The upper (blue) plot tracks the parameter $\gamma$. The optimizer systematically drives $\gamma$ downwards from its initial positive value to a stable negative value. By design, a negative $\gamma$ deterministically sets the gate parameter $\theta$ to zero. This action effectively turns off the controlled phase shift gate within the quantum circuit, which indicates the database has a $\mathbb{Z}_2\times \mathbb{Z}_2$ symmetry. 

By autonomously discovering the hidden symmetry within the database, our quantum photonic autoencoder has achieved efficient information compression.

{{\bf \em Less restricted ansatz for Simon's symmetry.---}}We also explore the possibility of using a more general circuit ansatz than the parametrised quantum Fourier transform in order to probe the limits of converging ability. We design a quantum circuit with two tunable parameters, as shown in Fig. \ref{fig:exp_up_all}(c). The parameters $\theta$ are associated with single qubit gates via
\begin{equation}
\label{eq:GofTheta}
G(\theta)\! = \!\cos(2\theta)\!\ket{0}\!\bra{0}+\sin(2\theta)\!(\ket{0}\!\bra{1}+\ket{1}\!\bra{0})-\cos(2\theta)\ket{1}\!\bra{1}\!.
\end{equation}
The top two qubits encode the IN register and the bottom qubit encode the OUT register. The unitary $U_{f'}$ transforms the state $|x\rangle_{IN}|0\rangle_{OUT}$ into $|x\rangle_{IN}|f'(x)\rangle_{OUT}$.

The hidden symmetry to be identified is set to be Simon's symmetry~\cite{simon1997power}. Thus the database function satisfies $f'(x)=f'(x')$ whenever $x'=x\oplus s$, where $s$ is an unknown bit string. 

The photonic implementation of the Fig. \ref{fig:exp_up_all}(c) circuit is depicted in Fig. \ref{fig:exp_up_all}(d). In our experimental setup, the IN qubits and OUT qubits are encoded in the path and time-bin degrees of freedom, respectively, of a single photon. Specifically, the four path modes, arranged from bottom to top, correspond to $\ket{00}$, $\ket{01}$, $\ket{10}$, and $\ket{11}$ for $s=01$, and to $\ket{00}$, $\ket{10}$, $\ket{01}$, and $\ket{11}$ for $s=10$. The outputs of the function are encoded in two distinct time-bins, while the polarization degree of freedom is exploited during intermediate processing.

Operations on the IN qubits are implemented by first converting the path-encoded states into polarization-encoded states using beam displacers (BDs). The desired unitary operations are then applied to the polarization qubits via electronically controlled half-wave plates (E-HWPs), and finally, the states are converted back to the path encoding. The overall operation executed by the E-HWPs on the IN qubits is given by $G(\theta)$ of Eq.\ref{eq:GofTheta}, where $\theta$ is the gate parameter determined by the setting angle of the E-HWP. 

The oracle unitary operation $U_{f'}$ is implemented by inserting a fixed HWP into the top two path modes. This insertion induces a temporal shift in these modes, setting their time-bin state to $\ket{1}$, while the time bin of the bottom two path modes encoded as $\ket{0}$. Thus, the unitary operation $U_{f'}\ket{x}_{IN}\ket{0}_{OUT}=\ket{x}_{IN}\ket{f'(x)}_{OUT}$, with $f'(x)=f'(x\oplus s)$, is achieved using the first encoding strategy when $s=01$, and the second encoding strategy when $s=10$.

Upon completion of training, the parametrized gate $G(\theta)$ converges to the Hadamard gate, thereby transforming the circuit in Fig. \ref{fig:exp_up_all}(c) into the quantum Fourier transform for a database with Simon's symmetry. This final circuit is identical to the one manually designed by Simon. Details regarding the experimental results and training method are provided in the Supplementary Material.

{{\bf\em Summary and outlook.---} We have designed a variational quantum autoencoder to compress databases with hidden subgroup symmetry and demonstrated its successful implementation in a photonic quantum system.  This illustrates how the quantum hidden subgroup algorithm can be adapted for information compression. This method is expected to be applicable to larger databases with wider symmetries \cite{liu2024information}.

Methods for scaling up the protocol include (i) introducing more degrees of freedom per photon, such as orbital angular momentum and different wavelengths, (ii)  increasing the dimensionality of each degree of freedom, such as adding more time bins and more spatial modes, (iii) increasing the number of photons, (iv) delegating more parts of the algorithm to the classical optimisation to further reduce the need for deep quantum circuits, (v) investigating potential vanishing gradient issues-the circuit ansatzes used here have a specific structure, so typicality arguments for random circuits suggesting vanishing gradients do not directly apply. 

{{\bf\em Acknowledgements.---}} OD acknowledges support from the National Natural Science Foundation of China (grant nos. 12050410246, 1200509, and 12050410245) and the City University of Hong Kong (Project No. 9610623). LZ acknowledges support from National Natural Science Foundation of China (grant nos. U24A2017, 12347104, and 12461160276), Innovation Program for Quantum Science and Technology (grant no. 2024ZD0300900), the National Key Research and Development Program of China (grant no. 2023YFC2205802), Natural Science Foundation of Jiangsu Province (grant nos. BK20243060 and BK20233001).

\bibliography{reference}

\end{document}